\documentclass[openacc]{rstransa}

\usepackage{graphicx}
\usepackage{txfonts}
\usepackage{color}
\newcommand{\apj}{Astrophys. J. }

\newcommand{\aap}{Astron. Astrophys.}
\newcommand{\apjl}{Astrophys. J. }
\newcommand{\mnras}{Mon. Not. R. Astron. Soc.}

\newcommand{\apjs}{Astrophys. J.}
\newcommand{\nat}{{\it Nature}}

\newcommand{\pasp}{Publ. Astron. Soc. Pac.}
\newcommand{\ssr}{SSR}

\newcommand{\revise}[1]{\textcolor{black}{#1}}
\titlehead{Research}

\begin{document}

\title{Strongly lensed supernovae: lessons learned}

\author{
A.Goobar, J. Johansson and A. Sagu\'es Carracedo
}

\address{The Oskar Klein Centre, Department of Physics, Stockholm University, SE 106 91 Stockholm, Sweden\\}

\subject{astrophysics, cosmology, time-domain astronomy}

\keywords{supernovae, gravitational lensing, time-domain surveys}

\corres{Ariel Goobar\\
\email{ariel@fysik.su.se}}

\begin{abstract}
\revise{Since about a decade}, we have finally entered the era of discoveries of multiply-imaged gravitationally lensed supernovae. To date, all cluster lensed supernovae, \revise{very distant, faint and spatially resolved}, have been found from space. \revise{In contrast,} those deflected by individual galaxies \revise{have been very compact and bright enough to be} identified with wide-field ground-based surveys through the magnification of "standard candles" method, i.e., without the need of spatially resolving the individual images. We review the challenges in identifying these extremely rare events, as well as the unique opportunities they offer for \revise{two major applications:} time-delay cosmography and the study of the properties of the deflecting bodies acting as lenses. 
\end{abstract}

\maketitle
\section{Introduction}
\revise{Successful applications for the use of supernova observations for precision cosmology became possible in the early 1990's when CCD cameras with scales exceeding a few arcminutes became available in astronomy.
"Scheduled" discoveries of high-redshift supernovae became possible \cite{1995ApJ...440L..41P,1997ApJ...483..565P}, which led to proposals to target observations towards massive lensing clusters, used as ``gravitational telescopes'' magnifying the flux from background sources, to search for the most distant supernovae  \cite{1988ApJ...335L...9K,2000MNRAS.319..549S,2003A&A...405..859G}.} 
The gain factor in exposure length is $\mu^2$, where $\mu$ is the flux amplification provided by the lens. However, this is partially balanced as the solid angle at the source planes shrinks by a factor $\mu$ behind the lens. Earlier attempts using ground-based optical and infrared instruments to do monthly cadenced observations of lensing clusters did not uncover any  multiply-imaged supernova \cite{2009A&A...507...71G,2011ApJ...742L...7A}, \revise{mainly because of the relatively shallow depth of the observations.}

\begin{figure}[b]
  \centering
  \includegraphics[width=1.0\linewidth]{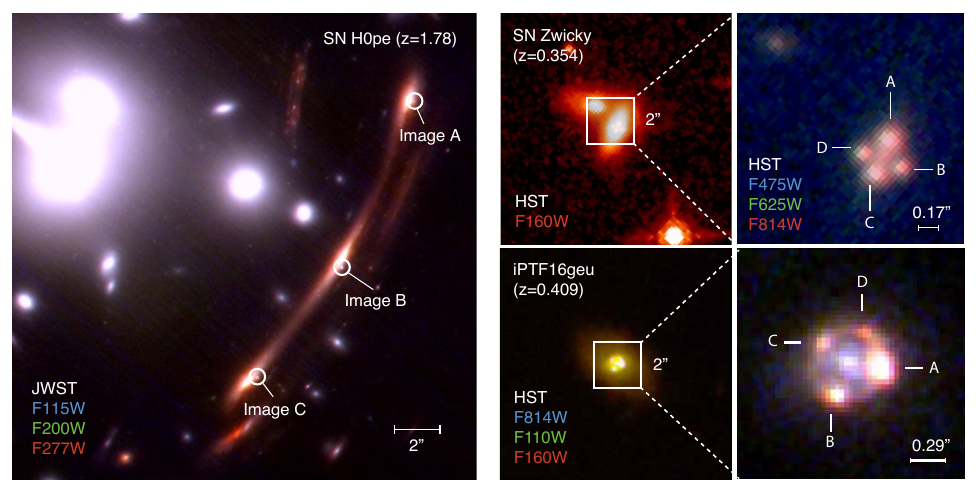}
  \caption{Gravitationally lensed Type Ia supernovae with multiple images, SN H0pe \cite{Frye+2023} ($z_s=1.78$; left panel) lensed by a cluster of galaxies and iPTF16geu \cite{2017Sci...356..291G} ($z_s= 0.409 $; bottom right panels) and SN Zwicky \cite{2023NatAs...7.1098G,2023ApJ...948..115P} ($z_s= 0.354$; top right panels) by individual galaxies. For the latter, the image flux ratios suggest that additional micro- or millilensing from  stellar objects or substructures is taking place in the deflecting galaxy \cite{2020MNRAS.496.3270M,2022A&A...662A..34D}. For iPTF16geu, a significant part of the intensity differences \revise{is} due to extinction in the lensing galaxy \cite{2020MNRAS.491.2639D}.} 
  \label{fig:lensedsn}
\end{figure}

It was only through \revise{observations with the Hubble Space Telescope (HST), which provided several magnitudes better sensitivity,} that the first cluster lensed supernova was found - a core-collapse supernova at $z_s=1.49$ lensed by the MACS J1149.6+2223 cluster \cite{2015Sci...347.1123K}. The supernova was named "SN Refsdal", honouring the memory of Sjur Refsdal who first proposed to use time delays between multiple images of \revise{gravitationally lensed supernova (glSNe)} to measure the Hubble constant \cite{1964MNRAS.128..307R}. Continued HST monitoring of massive clusters, and more recently also with JWST, has led to the discovery of several other multiply imaged supernovae behind clusters \cite{Rodney+2021,Chen+2022,Frye+2023,2024arXiv240402139P}, including three Type Ia supernovae (SNe Ia): SN H0pe and two "siblings", i.e., SNe hosted by the same galaxy, SN Requiem and SN Encore. While cluster lensed SNe have so far only been discovered from space, the three galaxy lensed SNe found to date were found in very wide-field surveys with ground-based telescopes. PS1-10afx was first reported as an unusual superluminous SN in the PanSTARRS transient survey \cite{2013ApJ...767..162C}. Three years after the SN discovery (too late for high-spatial resolution follow-up), it was shown in \cite{2013ApJ...768L..20Q} to be a highly magnified SN Ia at redshift $z_s = 1.388$, and \revise{with further observations} the lens at $z_l = 1.117$ was identified \cite{Quimby+2014}. 
Since then, two multiply-imaged SNe Ia have been found at Palomar Observatory: 
iPTF16geu, a SN Ia at redshift $z_s=0.409$, deflected by a galaxy at $z_l=0.2163$,  detected by the intermediate Palomar Transient Factory \cite{2017Sci...356..291G}; and another SNIa, SN Zwicky ($z_s=0.354; z_l=0.226$) \cite{2023NatAs...7.1098G} by the ongoing Zwicky Transient Facility (ZTF). Figure \ref{fig:lensedsn} shows \revise{space-based} imaging of three multiply-imaged SNe Ia, highlighting the different angular scales for cluster and galaxy lens systems found to date. 
For a recent review of the status of lensing of supernovae, see \cite{2024SSRv..220...13S}. The current manuscript focuses on lessons learned on gLSNe findings from ground-based transient surveys, iPTF16geu and SN Zwicky in particular \revise{and provides some discussion about implications for future surveys. }

\section{Spatially unresolved strongly lensed SNe}
\begin{figure}
  \centering
  \includegraphics[width=1.0\linewidth]{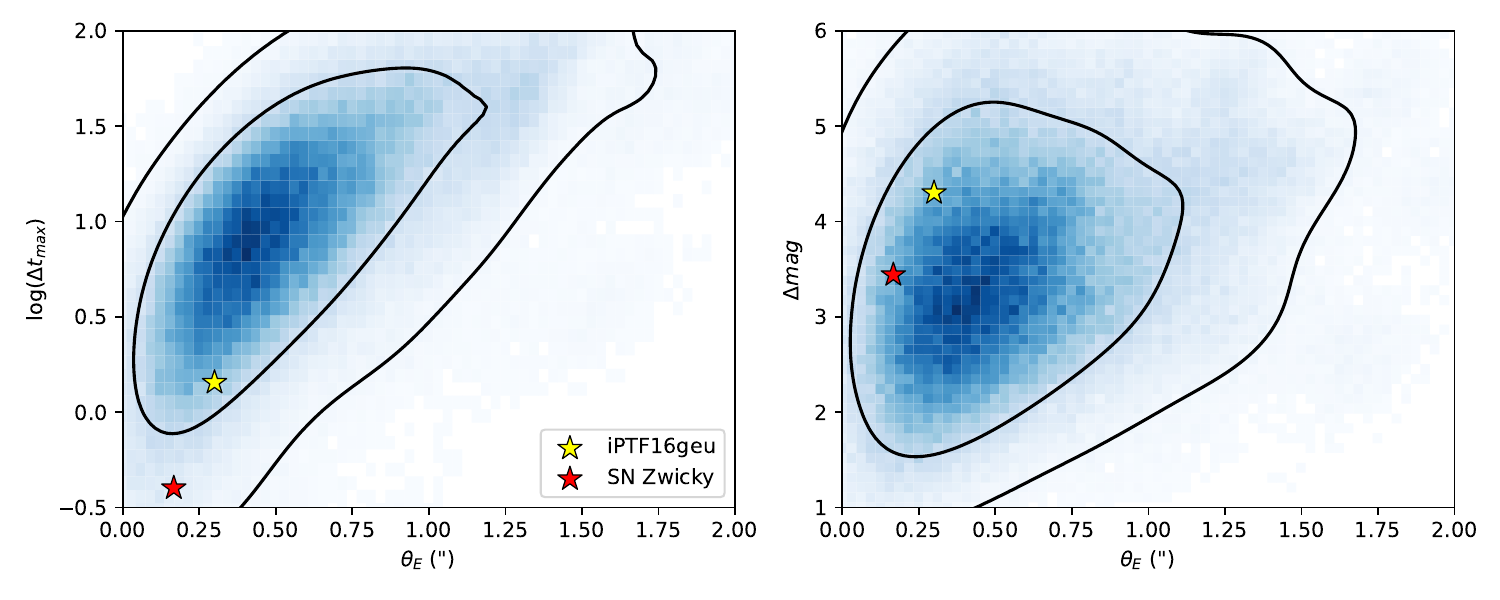}
  \caption{{\em Left:} Probability distribution of the time delays (in days) between multiple SN images vs the Einstein radius (arcseconds) for glSN systems expected in ZTF. {\em Right}: Magnification (in magnitudes) vs Einstein radius. The two stars, \revise{yellow and red}, show the location in the parameter space of iPTF16geu \revise{(yellow)} and SN Zwicky \revise{(red)}. The black contours indicate the 68\% and 95\% confidence regions.}
  \label{fig:separation}
\end{figure}

Wide-field imaging transient surveys like Palomar's PTF (2009-2012), iPTF (2013-2017) and ZTF (operating since 2018) have the ability to cover the entire visible sky from the Northern hemisphere in a single night. The extremely large search area of ZTF, facilitated by its 47.sq.~deg field-of-view camera, makes it particularly suitable for detecting rare transient phenomena. The limiting factors are the small collecting area of the 1.2m telescope and the very coarse spatial resolution. With 1$"$ pixel plate-scale and typically 2$"$ seeing at Palomar, detecting spatially resolved glSNe would be extremely rare. Simulations of the ZTF survey \cite{2019ApJS..243....6G} indicate that only about 2\% of the glSNe within discovery range from ZTF would have image separations exceeding 3$"$, at which point they could be detected as two individual point sources. 
The time delay between multiple images is typically shorter than the typical time scale of the lightcurves, \revise{making it quite} challenging to identify glSNe from the vast pool of regular SN lightcurves through multiple detections separated in time. Figure \ref{fig:separation} shows the distribution of time delays between SN images and the characteristic angular scale of strong lensing, the Einstein radius $\theta_E$ expected from simulations of the ZTF survey \cite{2024arXiv240600052S}. The extremely compact multi-image systems iPTF16geu $\theta_E=0.3"$ and SN Zwicky $\theta_E=0.16"$ (shown in Figure~\ref{fig:lensedsn}) \revise{were} identified through the magnification method. As indicated by stars in Figure \ref{fig:separation}, these two SNe where highly magnified $\Delta m = 2.5\log_{10}(\mu)>3$mag, split into four images. Thanks to the "standard candle" nature of Type Ia supernovae, \revise{which typically show a scatter of $\sim 0.15$ mag after applying corrections for lightcurve stretch and colour, they} can be easily identified as outliers in brightness, {\em provided a spectrum is available with the spectral classification and the redshift of the SN.} We will return to this issue in Section \ref{bottleneck}. \revise{The alternative path, searching for lensed SNe in already known lensing systems is very interesting, but is currently limited by the low statistics of known lenses. Furthermore, some of the lensed galaxy sources are likely to be too high redshifts, hence practically, supernovae would only be observable with very sensitive near-infrared instruments, i.e., basically only from space.   Hence, the situation should be much improved thanks to wide-field high-spatial resolution space imaging of the sky, starting with the Euclid mission \cite{2024arXiv240806217A}.}
\section{A different population of lens systems}
As can be appreciated from Figures \ref{fig:lensedsn} and \ref{fig:separation}, the systems found to date form the ground are extremely compact. Such small angular separation lensing systems are rarely found by other means given the extreme spatial resolution needed. Hence, it was shown in \cite{2023NatAs...7.1098G} that the glSNe uncover a \revise{previously} unexplored population of low stellar mass lensing galaxies. In particular, the compact systems provide interesting insights into the inner $\sim$1 kpc region of galaxies. The downside, discussed further in the Section \ref{timedelays}, is that they are not suitable for time-delay measurements.

\section{The discovery bottleneck: spectroscopic follow-up}
\label{bottleneck}
\begin{figure}
  \centering
  \includegraphics[width=0.8\linewidth]{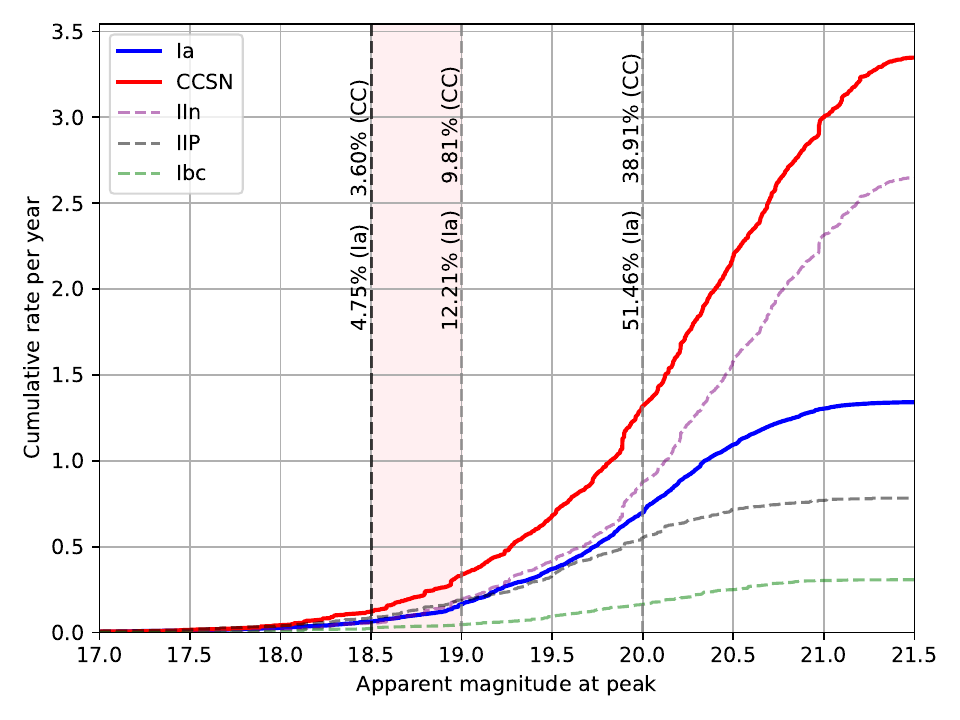}
  \caption{Expected yearly discovery \revise{rates by the ZTF survey (including detection efficiencies and weather losses)} as a function of \revise{apparent g and r-band (whichever is the brightest)} magnitude threshold for \revise{lensed SNe Ia (blue) and core-collapse supernovae (CC SNe; red)}. The dashed, fainter curves \revise{represent subtypes that are individual components of the red curve for CC SNe: IIP, IIn, and Ibc, with IIn being the dominant type}, assuming standard luminosity functions and constant fraction of CC  population, independent of redshift. The vertical dashed lines indicate the magnitude cuts of 18.5, 19, and 20 mag with the corresponding percentage of each SN type up to that cut. The 18.5 mag cut corresponds to the BTS magnitude completeness threshold, under which most supernovae are spectroscopically classified. We also include a region up to 19 mag as BTS extends to such magnitudes when the schedule allows it. Adapted from \cite{2024arXiv240600052S}.}
  \label{fig:rates}
\end{figure}

\revise{Due to} limited spectroscopic resources, only a small fraction of the transient discoveries in iPTF and ZTF were followed up with the necessary spectroscopic screening needed to identify a lensed SN. While the photometric detection threshold in ZTF is around 20.5-21 mag \cite{Bellm+2019}, the spectroscopic classification, as a part of the Bright Transient Survey (BTS) is only complete to 18.5 mag \cite{2020ApJ...895...32F}. Simulations of the ZTF survey \cite{2024arXiv240600052S} 
(see also \cite{2023MNRAS.526.4296S}) show that the bright threshold of the BTS spectroscopic classification has been the bottleneck for identifying the glSNe with the magnification method. 
In a recent \revise{study \cite{2024arXiv240518589T}, an archival analysis of ZTF data was conducted} to search for possible missed "live" candidates due to the magnitude limitation from BTS. The search efficiency was enhanced by having access to galaxy redshifts from the Dark Energy Spectroscopic Instrument (DESI), spatially associated with longlived, red candidates. The search has produced a handful of intriguing candidates. While superluminous supernovae (SLSNe) cannot be fully rejected as a possible explanation, two archival ZTF events, are significantly different from typical SLSNe and their lightcurves can be modelled as two-image lensed SNIa systems. From this two-image modelling, time delays of $22 \pm 3$ and 
$34 \pm 1$ days were estimated, respectively. If confirmed, \revise{e.g., if galaxy arcs were to be resolved upon future space observations, it suggests that we may have found the first events time delays measured with better than 15\% precision with ground-based resources!} 
The findings are in good agreement with the rate expectations from survey simulations in \cite{2024arXiv240600052S}, shown in Figure \ref{fig:rates}.
\section{Time delays and the quest for the Hubble constant}
\label{timedelays}
One of the main motivations behind searching for glSNe is to use their lightcurves to measure the time delays between the multiple images, from which the Hubble constant (H$_0$) can be inferred, as first suggested by Refsdal in 1964 \cite{1964MNRAS.128..307R}. In recent years, the interest in this type of measurement has gained a lot of \revise{attention} due to the emergence of the so called "Hubble tension". The value of H$_0$ obtained from the early universe \revise{cosmic microwave background (CMB)} anisotropy data, extrapolated to the present \revise{time} using the \revise{standard model of cosmology, $\Lambda$CDM,}  ($67.4 \pm 0.5$ km s$^{-1}$Mpc$^{-1}$), is in conflict with the local distance ladder measurement from the \revise{"Supernova H$_0$ for the Equation of State" (SH0ES)} team 
($73.0 \pm 1.0$ km s$^{-1}$Mpc$^{-1}$), see \cite{2023ARNPS..73..153K} for a recent review. Time-delay cosmography offers an interesting independent way to measure the Hubble constant and could provide further support or reject the notion that physics beyond the $\Lambda$CDM model is required. \revise{This is particularly interesting since we do not yet know what cold dark matter (CDM) is made of, nor if Einstein's cosmological constant $\Lambda$ is indeed what is causing the accelerated expansion of the universe.}

For many years, time-delay cosmography were carried out exclusively with multiply-imaged quasars \revise{(QSOs)}. \revise{While these studies have been extremely exciting,} the results \revise{remain} inconclusive (see e.g. \cite{2022arXiv221010833B} for a review). The smooth lightcurves of supernovae coupled with their favourable time scales make them potentially superior to QSOs for time-delay cosmography, \revise{given that QSO monitoring requires many years, and sometimes decades}. Furthermore, unlike QSOs, supernovae fade in roughly a year time-scale, and ease significantly the modeling of the lens without contamination from the lensed images. Hence, the possibility to complement the
time-delay cosmography from QSOs with glSNe has generated a lot of interest. 

Thanks to the standard-candle nature of SNe Ia (after corrections for colour and lightcurve shape), their magnification  can be inferred up to an uncertainty related to their intrinsic luminosity scatter, about 0.15 mag. This is potentially a key feature, since it
can be used to break the so-called mass-sheet degeneracy \revise{in galaxy-scale lens systems}. In brief, the presence of a constant sheet of surface mass density leaves the predicted images unchanged, but alters the time delay between the images \cite{Falco+1985}. Breaking the mass-sheet degeneracy, e.g., through the
model independent measurement of the SNIa magnification, is therefore a very important element for constraining H$_0$ \cite{Birrer+2021}.

In the following sections we will discuss some additional challenges to break the mass-sheet degeneracy posed by extinction by dust in the host and deflecting galaxy, as well as micro- and millilensing.

\subsection{Cluster scale lenses}
Through the monitoring of the multiple images of SN Refsdal \cite{Rodney+2016} the time delays and magnification ratios among the images were measured, and the most accurate time delay between  a pair of images was $376.0^{+5.6}_{-5.5}$\,days \cite{Kelly+2023a}.  This time-delay measurement with a relative uncertainty of 1.5\% provided the first \revise{and so far most} precise $H_0$ measurement from lensed SNe. 
\revise{Using two independent lensing models of the cluster, \cite{Kelly+2023a} found H$_0 = 66.6^{+4.1}_{-3.3}$ km s$^{-1}$Mpc$^{-1}$, while a separate study \cite{2024A&A...684L..23G} arrived at a similar result, H$_0=65.1^{+3.5}_{-3.4}$ km s$^{-1}$Mpc$^{-1}$.}

 More recently, $H_0$ was measured for SN H0pe. A combination of a spectroscopic \revise{(see Section~\ref{specdelay}) and photometric time-delay measurements \cite{2024arXiv240319029C,2024arXiv240318954P}} were compared to the predictions of many cluster lens models to measure a value for the Hubble constant \cite{2024arXiv240318902P}. Combined with the magnification of this SN~Ia, this yielded a value of $H_0 = 75.4^{+8.1}_{-5.5}$  km s$^{-1}$Mpc$^{-1}$.
 
\subsection{Galaxy scale lenses}
\begin{figure}
  \centering
  \includegraphics[width=0.8\linewidth]{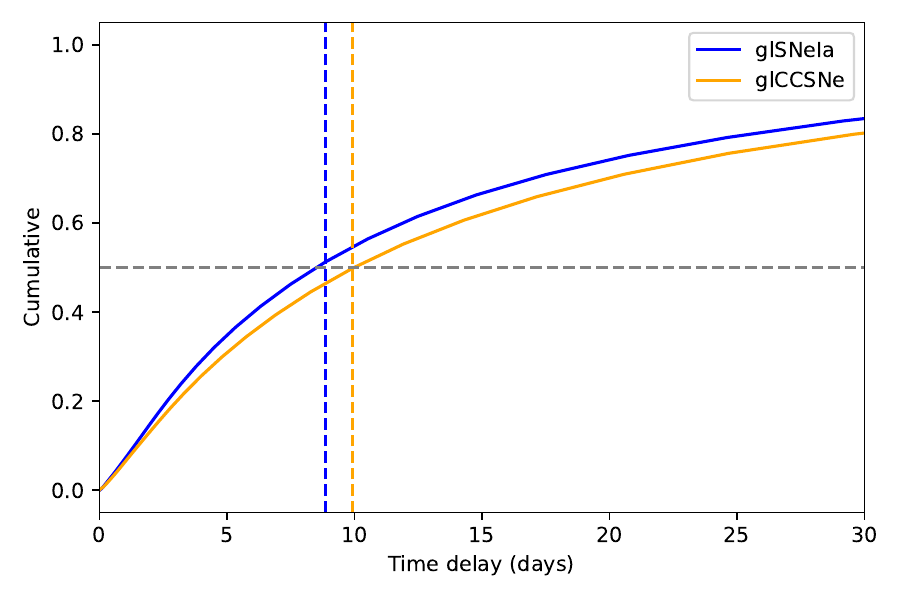}
  \caption{Expected cumulative distribution of time delays for glSNe discovered by ZTF, \revise{using all public and proprietary data in three filters,} showing that the median time delay is close to 10 days for both core-collapse and Type Ia supernovae. \revise{It should be noted that, unlike for SNe Ia, the luminosity functions and relative rates of CCSNe are poorly quantified. The somewhat longer time delays predicted by the simulations for lensed CCSNe is driven by the assumed intrinsically very bright IIn population, where considerable uncertainties remain.}
  Adapted from \cite{2024arXiv240600052S}.}
  \label{fig:delta_t}
\end{figure}

While the $H_0$ measurements from SN Refsdal and SN H0pe are very encouraging and exciting, some caution needs to be 
exercised in interpreting these results. The lensing models are very challenging since the mass distributions of clusters are quite complex, implying that the multiple image region of clusters is expected to be rich in substructures. For that reason, galaxy lenses are much simpler to model and therefore preferable, as they involve smaller systematic uncertainties.
However, the highly magnified compact systems within reach for shallow surveys like ZTF are expected to produce images which can be separated by just a few days, as shown in Figure \ref{fig:delta_t}, making it rather challenging for \revise{precise} measurements of time delays. That was the case for both iPTF16geu \cite{2020MNRAS.491.2639D}
and SN Zwicky \cite{2023NatAs...7.1098G}, as outlined in the next Section.

\subsection{The second maximum in the SN Ia near-IR lightcurves}
\label{sec:second_bump}
\begin{figure}
  \centering
  \includegraphics[width=1.0\linewidth]{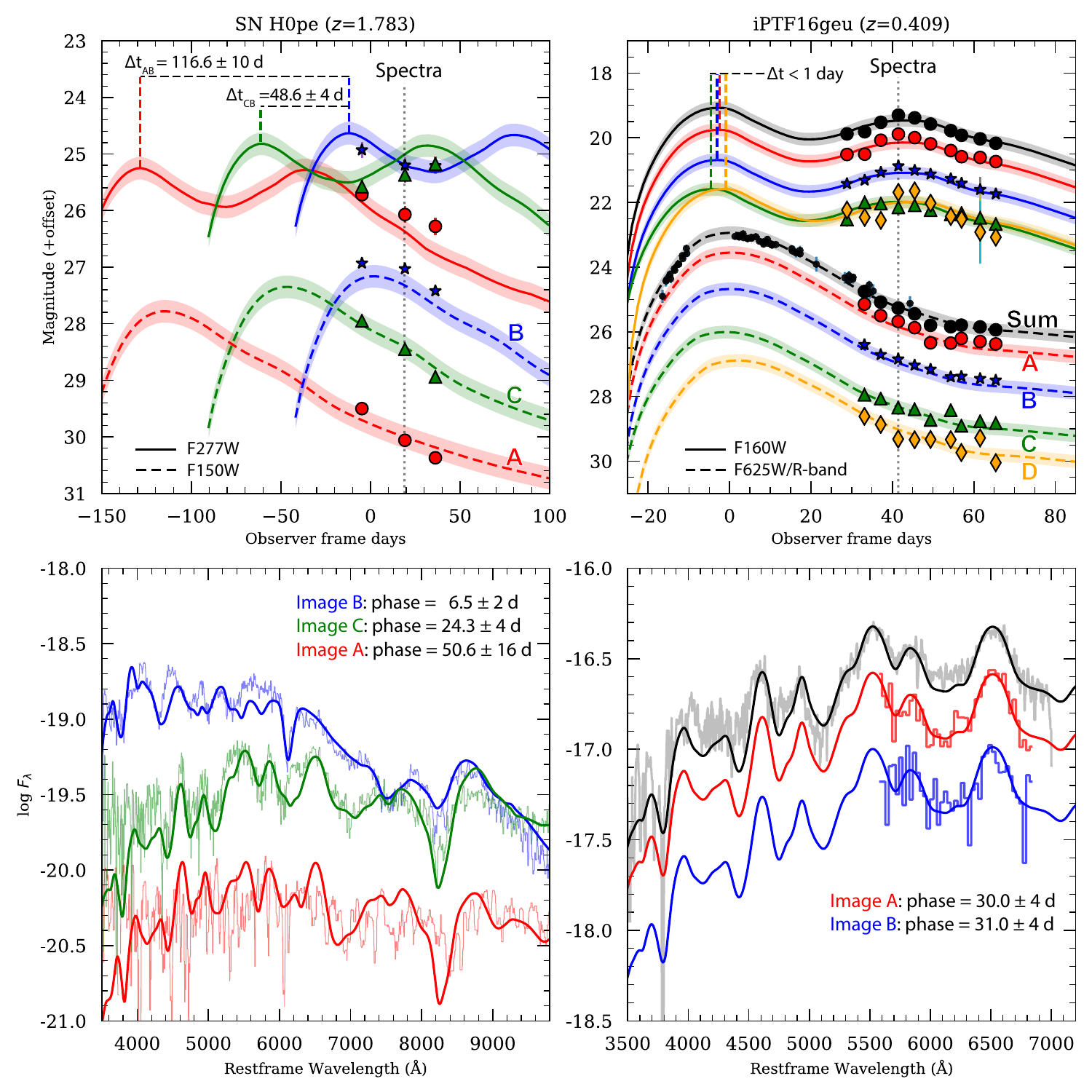}
  \caption{Top row: Photometric time-delay estimates from resolved images of SN H0pe \revise{from JWST} (left) and iPTF16geu \revise{from HST} (right). In both cases, the "second IR maximum" was used to measure the difference in arrival times between the multiple SN images. Bottom row: spectroscopic time delays \revise{for SN H0PE (left, JWST) and iPTF16geu (right, HST)} where the spectral features are dated using the SED template in \cite{hsiao2007}. Observations and further details can be found in
  \cite{2024arXiv240318954P,2024arXiv240319029C,2020MNRAS.491.2639D,2020MNRAS.496.3270M}.} 
  \label{fig:time-delay}
\end{figure}

Besides their "standard candle" nature, SNe Ia offer additional benefits for time-delay cosmography.  
\revise{While the optical bands for SNe Ia, like most other SNe, show a single peak, the SN~Ia near-IR bands display two peaks. It has been suggested that the secondary maximum originates from  the ionisation transition of iron-group elements in the ejecta, shifting from doubly- to singly-ionised as the temperature drops below about 7000 K \cite{2006ApJ...649..939K,2014MNRAS.441.3249D,2015MNRAS.449.3581J}.
  For the restframe lightcurves in bands beyond the $r$ filter, the secondary maximum appears within about a month after the restframe $B$-band lightcurve peak and can be used to measure photometric time delays.} 
  This is extremely useful \revise{because it allows for accurate measurements of arrival time differences between SN images, even if} the first maximum is missed or poorly sampled, as was the case for both iPTF16geu and SN H0pe, shown in Figure \ref{fig:time-delay}.

\subsection{Spectroscopic time delays}
\label{specdelay}
The two bottom panels of Figure \ref{fig:time-delay} \revise{illustrate another unique advantage of using} supernovae for time-delay cosmography. During the early phases of the supernova explosion, the atmosphere is thick and the supernova spectral energy distribution is formed by the  outer layers with lower opacity, \revise{known as the photosphere}. For a homologous expansion, $r= v\cdot t$, that corresponds to very high velocities. As the expansion thins out the atmosphere, the photosphere recedes, and the typical absorption features come from closer the centre, hence \revise{at} lower velocities. This change of velocities of the SN features can be used to extract the phase of the supernova at the time of observations. Spectroscopic time-delay measurements have been carried out successfully for iPTF16geu \cite{Johansson+2021} and SN H0pe \cite{2024arXiv240319029C}, as shown in Figure \ref{fig:time-delay}.

\subsection{Measuring time delays with (mostly) unresolved data}
\begin{figure}
  \centering
  \includegraphics[width=1.0\linewidth]{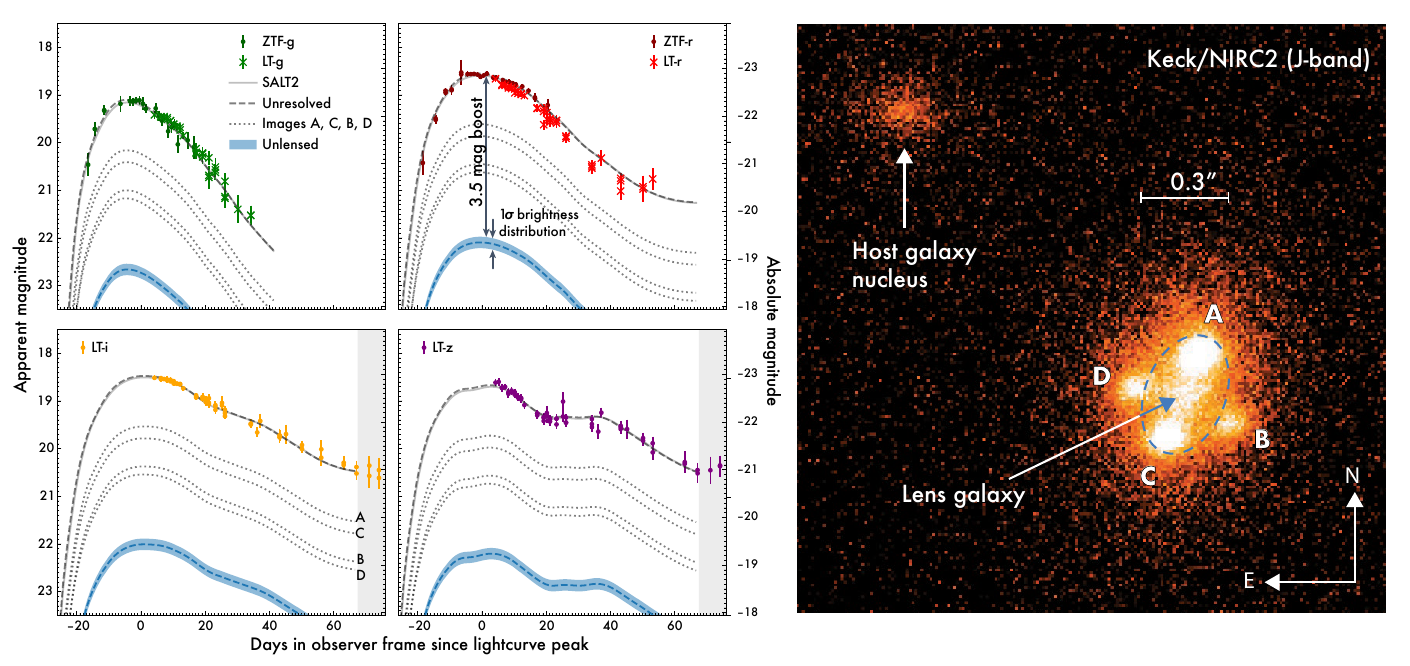}
  \caption{Ground-based (unresolved) lightcurves for SN Zwicky and resolved Keck/AO images from \cite{2023NatAs...7.1098G}. Using the prior from the Keck/AO image showing the quad configuartion and the fluxes at one epoch, time delays could be measured accurately with the unresolved multi-band data.}
  \label{fig:unresolved}
\end{figure}

One of the important recent developments is the realisation that time-delays can be inferred from unresolved lightcurves, provided there is at least one high-spatial resolution image of the system that gives the image multiplicity, their positions, and the relative image fluxes, as shown in Figure \ref{fig:unresolved} from \cite{2023NatAs...7.1098G}. 
\revise{This is potentially very important as obtaining well-sampled lightcurves from space or adaptive optics instruments may be prohibitively demanding for a sample of glSNe.}
\revise{Hence, the positive outcome of the fits to the unresolved images of SN Zwicky deserve some attention.} The publicly available, python-based software \texttt{sntd} \cite{Pierel2019} was used for inferring the restframe $B$-peak magnitude, the lightcurve shape and colour SN~Ia SALT2 parameters \cite{Guy:2007js} and the time-delays between the images.
Unresolved photometry from the Palomar and Liverpool telescopes in $g,r,i,z$ filters were included in the fit, along with a model including the flux contributions from the four sets of lightcurves accounting for extinction, each with their own time of maximum. The fit is constrained by imposing a prior on the image ratios at the date of the Keck/NIRC2 observations, shown in the right-hand side panel of Figure~\ref{fig:unresolved}. The total lensing magnification was fitted to be $\mu = 24.3 \pm 2.7$. Although negligible time delays were found in this case, the method is very promising for future systems.

\section{The interstellar medium and differential extinction}\label{dust}
\begin{figure}
  \centering
  \includegraphics[width=0.6\linewidth]{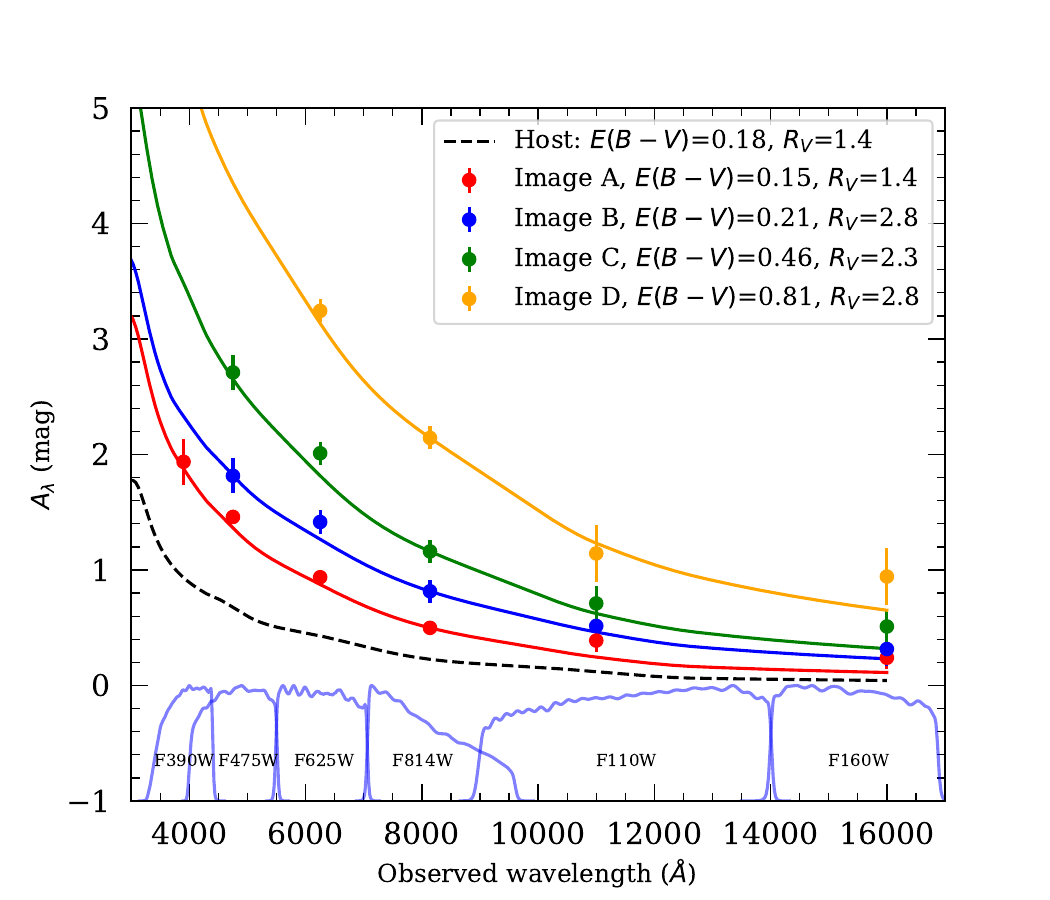}
  \caption{Inferred wavelength dependent extinction, $A_\lambda$, for the four resolved images of iPTF16geu measured with HST.
    The absorption from the host galaxy dust is plotted with dotted black line. For Image A we can see that the host galaxy is the dominant source of extinction, \revise{while} for Images B, C, D there is a progressively larger contribution from the dust in the lens galaxy (see \cite{2020MNRAS.491.2639D} for data and analysis information).} 
  \label{fig:dust}
\end{figure}

As the light of gLSNe pierces through the inner regions of the deflecting galaxies, measurements of magnification crucially depend on the ability to accurately correct for losses due to scattering on dust grains, both in the host and lensing galaxy. For that purpose, \revise{spatially resolved} multi-band imaging is used, since the magnitude increase (i.e., loss \revise{of} flux in logarithmic units) due to dust extinction is roughly inversely proportional \revise{to} wavelength, $A_\lambda \propto \lambda^{-1}$. To \revise{complicate matters further}, observational evidence suggests that the composition and grain size distribution in the extragalactic interstellar medium could be very diverse \cite{2015MNRAS.453.3300A}, with total-to-selective extinction $R_V = A_V/E(B-V)$ potentially quite different from the Milky-Way value, hence the need to both fit the colour excess $E(B-V)$ and 
$R_V$, even from individual images. Since precise information of the range of properties for dimming by dust in other galaxies is so critical for accurate distance measurements in cosmology, it is very exciting to be able to carry out such measurements with resolved images of lensed SNe, as was the case for iPTF16geu \revise{(originally discovered in unresolved observations from the ground)} \cite{2020MNRAS.491.2639D}, shown in Figure \ref{fig:dust}. Images C and D (see Figure \ref{fig:lensedsn}) \revise{were} particularly reddened, and thanks to having HST images in at least four filters useful constraints could be set on both the colour excess and $R_V$, providing unique a test of dust grain density and properties \revise{along} multiple lines of sight in an intermediate redshift galaxy.  
\revise{In terms of suitable follow-up strategy of resolved images, we can see that at least three filters are needed to fit the two extinction parameters, $E(B-V)$ and $R_V$. In addition, as described in Sec.~\ref{sec:second_bump}, restframe near-infrared observations allow to measure time delays past maximum. In conclusion, four filters constitute a good minimum set of filters to study multiple images of glSNe.}

\section{Macro vs milli/microlensing}

Using multi-band follow-up observations of iPTF16geu with HST, an accurate (model independent) measurement of the total magnification was made, $\mu=67.8^{+2.6}_{-2.9}$ \cite{2020MNRAS.491.2639D}, after correction for non-negligible extinction by dust in both the host and lens galaxies, as  discussed in Section \ref{dust}. The time delays between the SN images for this system were very small, about a day or less \cite{More+2017,2020MNRAS.491.2639D}.  However, the flux ratios between the supernova images (see Fig.~\ref{fig:lensedsn}) were not consistent with expectations from  (macro) lensing of a smooth extended deflector, even accounting for differential extinction, hinting at additional lensing contributions from galactic sub halos (millilensing) or stellar objects (microlensing) \cite{2020MNRAS.496.3270M}. In either case, the sub-spliting of the SN images is too small to be resolved as it is of order milli-arcseconds or less, whereas e.g., the HST angular resolution is at least a few tens of milli-arcseconds. The situation was very similar for SN Zwicky ($\mu=23.7 \pm 3.2$), except that there was no ambiguity between dimming by dust and microlensing (de-)magnification \cite{2023NatAs...7.1098G,2023ApJ...948..115P}. While extinction by dust can be identified and corrected through its wavelength dependence, microlensing by stellar objects is a more severe challenge \cite{DoblerKeeton2006}. The intrinsic size of a SN is comparable to the Einstein radius of an individual star in the deflecting galaxy. Hence, the observed magnification is sensitive to the  unknown positions of stars and substructures in the lensing galaxy. A thorough discussion on microlensing of SNe can be found in \cite{2024SSRv..220...13S} and potential means to mitigate this issue has been discussed in e.g., \cite{Foxley-Marrable+2018,2024arXiv240303264W}. On the positive side, the image flux ratios observed can be used to infer limits on the possible dark matter contribution from e.g., primordial black holes over a wide mass range \cite{2022A&A...662A..34D}.

\section{glSNe in the LSST era}
Thanks to the wide-field coverage and faint photometric limit of the Legacy Survey of Space and Time (LSST) at the Vera Rubin Observatory, the discovery rate of galaxy-scale lenses, as well as strongly lensed supernovae is expected to increase dramatically. \revise{For a general overview of strong lensing by galaxies in LSST, see \cite{2024arXiv240608919S}}.
Simulation studies have shown that hundreds of lensed SNe should be found \cite{2019ApJS..243....6G}, many of which will be spatially resolved \cite{Wojtak+2019}. 
\revise{For glSNe unresolved in LSST, the situation may be rather challenging, given the difficulties of identifying these with scarce spectroscopic follow-up, as discussed in Section \ref{bottleneck}. Unlike the case for ZTF glSNe, lensed SNe in LSST would occur at "normal" redshifts for unlensed supernovae, hence adding to the difficulties in "cherry-picking" candidates for high spatial-resolution observations. In some cases, with good enough cadence and signal-to-noise ratio, multiple peaks in the lightcurves corresponding to multiple images should be detectable.
An additional interesting path to explore is the ability to use LSST galaxy multi-band imaging to identify overlapping galaxies, for which the measured colours would be incompatible with originating from a galaxy at a single redshift, hence potentially a good way to identify galaxy-galaxy lensed systems\footnote{This is being pursued by Peter Nugent and collaborators, private communication.}. For about half of the glSNe to be found with LSST, the host galaxy will also be in the area of strong lensing \cite{2024arXiv240704080S}. Furthermore, as suggested by the authors of \cite{2024arXiv240704080S}, these systems may be spatially resolved with observations from the Euclid space mission, or eventually with the Roman Space Telescope} 

\revise{
Even when two SN images are found, the contamination from random coincidences of two independent SNe in the same galaxy, known as "siblings", cannot be dismissed. For the ZTF survey, these are much more common than glSNe \cite{2024arXiv240601434D}, and e.g., \cite{2020ApJ...896L..13S} estimate that about 800 siblings where both SNe are of Type Ia will be found by LSST. Since the volumetric rate of CCSNe exceeds that for  SNe Ia, especially at high redshift, many thousands of sibling systems should be present in the LSST data. While the majority could be easily rejected because of being mixed SN types, too large angular or temporal separation, the false alarm rate will not be negligible. In addition to siblings, there will be many cases where two SNe occur in different galaxies, with small angular separation. Clearly, significant efforts and observational resources will be needed to deliver precision cosmology with glSNe, also in the LSST era. However, the effort should be worthwhile.} For \revise{resolved} SNe Ia, \cite{2024MNRAS.531.3509A} found that a `gold sample' of  $\sim$ 10 lensed SNIa per year can be expected, with time delays exceeding 10 days, caught before peak, and sufficiently bright (below 22.5 mag) for spectroscopic follow-up observations. In three years of LSST operations, such a sample can yield a 1.5\% measurement of the Hubble constant.

\section{Conclusion}
The rapid developments in time-domain astronomy, including wide-field imaging from the ground and very sensitive near-IR space instruments has led to an exciting development in the discovery of strongly lensed supernovae: the era of time-delay cosmography with supernovae has begun! Ground-based searches have uncovered a population of compact lens systems, where micro and millilensing effects are very important to characterise, also since they provide tests for the nature of dark matter. Lessons learned from searches to date give rise to great optimism, as instruments to be deployed in the immediate future will greatly enhance the feasibility and science reach of cosmology and astrophysics with lensed supernovae.   

\vskip6pt

\ack{Many thanks to the Royal Society for the great hospitality during the meeting. The authors acknowledge the work behind these results from whole the ZTF lensed supernova working group as well as financial support from {\em Vetenskapsr\aa det}, the Swedish Research Council through grants project Dnr 2020-03444, the G.R.E.A.T research environment, Dnr 2016-06012, and the Swedish National Space Agency, Dnr 2023-00226. \revise{Excellent comments from two reviewers helped to improve the manuscript.}}



\begin{thebibliography}{99}

\bibitem{1995ApJ...440L..41P}
{Perlmutter} S, {Pennypacker} CR, {Goldhaber} G, {Goobar} A, {Muller} RA,
  {Newberg} HJM, {Desai} J, {Kim} AG, {Kim} MY, {Small} IA, {Boyle} BJ,
  {Crawford} CS, {McMahon} RG, {Bunclark} PS, {Carter} D, {Irwin} MJ,
  {Terlevich} RJ, {Ellis} RS, {Glazebrook} K, {Couch} WJ, {Mould} JR, {Small}
  TA, {Abraham} RG. 1995  {A Supernova at Z = 0.458 and Implications for
  Measuring the Cosmological Deceleration}. {\em \apjl} \textbf{440}, L41.
(\href{http://dx.doi.org/10.1086/187756}{10.1086/187756})

\bibitem{1997ApJ...483..565P}
{Perlmutter} S, {Gabi} S, {Goldhaber} G, {Goobar} A, {Groom} DE, {Hook} IM,
  {Kim} AG, {Kim} MY, {Lee} JC, {Pain} R, {Pennypacker} CR, {Small} IA, {Ellis}
  RS, {McMahon} RG, {Boyle} BJ, {Bunclark} PS, {Carter} D, {Irwin} MJ,
  {Glazebrook} K, {Newberg} HJM, {Filippenko} AV, {Matheson} T, {Dopita} M,
  {Couch} WJ. 1997  {Measurements of the Cosmological Parameters
  {\ensuremath{\Omega}} and {\ensuremath{\Lambda}} from the First Seven
  Supernovae at z >= 0.35}. {\em \apj} \textbf{483}, 565--581.
(\href{http://dx.doi.org/10.1086/304265}{10.1086/304265})

\bibitem{1988ApJ...335L...9K}
{Kovner} I, {Paczynski} B. 1988  {Supernovae in Luminous Arcs}. {\em \apjl}
  \textbf{335}, L9.
(\href{http://dx.doi.org/10.1086/185328}{10.1086/185328})

\bibitem{2000MNRAS.319..549S}
{Sullivan} M, {Ellis} R, {Nugent} P, {Smail} I, {Madau} P. 2000  {A strategy
  for finding gravitationally lensed distant supernovae}. {\em \mnras}
  \textbf{319}, 549--556.
(\href{http://dx.doi.org/10.1046/j.1365-8711.2000.03875.x}{10.1046/j.1365-8711.2000.03875.x})

\bibitem{2003A&A...405..859G}
{Gunnarsson} C, {Goobar} A. 2003  {Massive galaxy clusters as gravitational
  telescopes for distant supernovae}. {\em \aap} \textbf{405}, 859--866.
(\href{http://dx.doi.org/10.1051/0004-6361:20030648}{10.1051/0004-6361:20030648})

\bibitem{2009A&A...507...71G}
{Goobar} A, {Paech} K, {Stanishev} V, {Amanullah} R, {Dahl{\'e}n} T,
  {J{\"o}nsson} J, {Kneib} JP, {Lidman} C, {Limousin} M, {M{\"o}rtsell} E,
  {Nobili} S, {Richard} J, {Riehm} T, {von Strauss} M. 2009  {Near-IR search
  for lensed supernovae behind galaxy clusters. II. First detection and future
  prospects}. {\em \aap} \textbf{507}, 71--83.
(\href{http://dx.doi.org/10.1051/0004-6361/200811254}{10.1051/0004-6361/200811254})

\bibitem{2011ApJ...742L...7A}
{Amanullah} R, {Goobar} A, {Cl{\'e}ment} B, {Cuby} JG, {Dahle} H, {Dahl{\'e}n}
  T, {Hjorth} J, {Fabbro} S, {J{\"o}nsson} J, {Kneib} JP, {Lidman} C,
  {Limousin} M, {Milvang-Jensen} B, {M{\"o}rtsell} E, {Nordin} J, {Paech} K,
  {Richard} J, {Riehm} T, {Stanishev} V, {Watson} D. 2011  {A Highly Magnified
  Supernova at z = 1.703 behind the Massive Galaxy Cluster A1689}. {\em \apjl}
  \textbf{742}, L7.
(\href{http://dx.doi.org/10.1088/2041-8205/742/1/L7}{10.1088/2041-8205/742/1/L7})

\bibitem{2015Sci...347.1123K}
{Kelly} PL, {Rodney} SA, {Treu} T, {Foley} RJ, {Brammer} G, {Schmidt} KB,
  {Zitrin} A, {Sonnenfeld} A, {Strolger} LG, {Graur} O, {Filippenko} AV, {Jha}
  SW, {Riess} AG, {Bradac} M, {Weiner} BJ, {Scolnic} D, {Malkan} MA, {von der
  Linden} A, {Trenti} M, {Hjorth} J, {Gavazzi} R, {Fontana} A, {Merten} JC,
  {McCully} C, {Jones} T, {Postman} M, {Dressler} A, {Patel} B, {Cenko} SB,
  {Graham} ML, {Tucker} BE. 2015  {Multiple images of a highly magnified
  supernova formed by an early-type cluster galaxy lens}. {\em Science}
  \textbf{347}, 1123--1126.
(\href{http://dx.doi.org/10.1126/science.aaa3350}{10.1126/science.aaa3350})

\bibitem{1964MNRAS.128..307R}
{Refsdal} S. 1964  {On the possibility of determining Hubble's parameter and
  the masses of galaxies from the gravitational lens effect}. {\em \mnras}
  \textbf{128}, 307.
(\href{http://dx.doi.org/10.1093/mnras/128.4.307}{10.1093/mnras/128.4.307})

\bibitem{Rodney+2021}
{Rodney} SA, {Brammer} GB, {Pierel} JDR, {Richard} J, {Toft} S, {O'Connor} KF,
  {Akhshik} M, {Whitaker} K. 2021  {A Gravitationally Lensed Supernova with an
  Observable Two-Decade Time Delay}. {\em arXiv e-prints} p. arXiv:2106.08935.

\bibitem{Chen+2022}
{Chen} W, {Kelly} PL, {Oguri} M, {Broadhurst} TJ, {Diego} JM, {Emami} N,
  {Filippenko} AV, {Treu} TL, {Zitrin} A. 2022  {Shock cooling of a
  red-supergiant supernova at redshift 3 in lensed images}. {\em \nat}
  \textbf{611}, 256--259.
(\href{http://dx.doi.org/10.1038/s41586-022-05252-5}{10.1038/s41586-022-05252-5})

\bibitem{Frye+2023}
{Frye} BL, {Pascale} M, {Pierel} J, {Chen} W, {Foo} N, {Leimbach} R, {Garuda}
  N, {Cohen} S, {Kamieneski} P, {Windhorst} R, {Koekemoer} AM, {Kelly} P,
  {Summers} J, {Engesser} M, {Liu} D, {Furtak} L, {Polletta} M, {Harrington} K,
  {Willner} S, {Diego} JM, {Jansen} R, {Coe} D, {Conselice} CJ, {Dai} L, {Dole}
  H, {D'Silva} JCJ, {Driver} S, {Grogin} N, {Marshall} MA, {Meena} A, {Nonino}
  M, {Ortiz}, Rafael I, {Pirzkal} N, {Robotham} A, {Ryan} RE, {Strolger} L,
  {Tompkins} S, {Trussler} J, {Willmer} C, {Yan} H, {Yun} MS, {Zitrin} A. 2023
  {The JWST Discovery of the Triply-imaged Type Ia ``Supernova H0pe'' and
  Observations of the Galaxy Cluster PLCK G165.7+67.0}. {\em arXiv e-prints} p.
  arXiv:2309.07326.
(\href{http://dx.doi.org/10.48550/arXiv.2309.07326}{10.48550/arXiv.2309.07326})

\bibitem{2024arXiv240402139P}
{Pierel} JDR, {Newman} AB, {Dhawan} S, {Gu} M, {Joshi} BA, {Li} T, {Schuldt} S,
  {Strolger} LG, {Suyu} SH, {Caminha} GB, {Cohen} SH, {Diego} JM, {Dsilva} JCJ,
  {Ertl} S, {Frye} BL, {Granata} G, {Grillo} C, {Koekemoer} AM, {Li} J,
  {Robotham} A, {Summers} J, {Treu} T, {Windhorst} RA, {Zitrin} A, {Agarwal} S,
  {Agrawal} A, {Arendse} N, {Belli} S, {Burns} C, {Ca{\~n}ameras} R,
  {Chakrabarti} S, {Chen} W, {Collett} TE, {Coulter} DA, {Ellis} RS, {Engesser}
  M, {Foo} N, {Fox} OD, {Gall} C, {Garuda} N, {Gezari} S, {Gomez} S,
  {Glazebrook} K, {Hjorth} J, {Huang} X, {Jha} SW, {Kamieneski} PS, {Kelly} P,
  {Larison} C, {Moustakas} LA, {Pascale} M, {P{\'e}rez-Fournon} I,
  {Petrushevska} T, {Poidevin} F, {Rest} A, {Shahbandeh} M, {Shajib} AJ,
  {Siebert} M, {Storfer} C, {Talbot} M, {Wang} Q, {Wevers} T, {Zenati} Y. 2024
  {Lensed Type Ia Supernova ``Encore'' at z=2: The First Instance of Two
  Multiply-Imaged Supernovae in the Same Host Galaxy}. {\em arXiv e-prints} p.
  arXiv:2404.02139.
(\href{http://dx.doi.org/10.48550/arXiv.2404.02139}{10.48550/arXiv.2404.02139})

\bibitem{2013ApJ...767..162C}
{Chornock} R, {Berger} E, {Rest} A, {Milisavljevic} D, {Lunnan} R, {Foley} RJ,
  {Soderberg} AM, {Smartt} SJ, {Burgasser} AJ, {Challis} P, {Chomiuk} L,
  {Czekala} I, {Drout} M, {Fong} W, {Huber} ME, {Kirshner} RP, {Leibler} C,
  {McLeod} B, {Marion} GH, {Narayan} G, {Riess} AG, {Roth} KC, {Sanders} NE,
  {Scolnic} D, {Smith} K, {Stubbs} CW, {Tonry} JL, {Valenti} S, {Burgett} WS,
  {Chambers} KC, {Hodapp} KW, {Kaiser} N, {Kudritzki} RP, {Magnier} EA, {Price}
  PA. 2013  {PS1-10afx at z = 1.388: Pan-STARRS1 Discovery of a New Type of
  Superluminous Supernova}. {\em \apj} \textbf{767}, 162.
(\href{http://dx.doi.org/10.1088/0004-637X/767/2/162}{10.1088/0004-637X/767/2/162})

\bibitem{2013ApJ...768L..20Q}
{Quimby} RM, {Werner} MC, {Oguri} M, {More} S, {More} A, {Tanaka} M, {Nomoto}
  K, {Moriya} TJ, {Folatelli} G, {Maeda} K, {Bersten} M. 2013  {Extraordinary
  Magnification of the Ordinary Type Ia Supernova PS1-10afx}. {\em \apjl}
  \textbf{768}, L20.
(\href{http://dx.doi.org/10.1088/2041-8205/768/1/L20}{10.1088/2041-8205/768/1/L20})

\bibitem{Quimby+2014}
{Quimby} RM, {Oguri} M, {More} A, {More} S, {Moriya} TJ, {Werner} MC, {Tanaka}
  M, {Folatelli} G, {Bersten} MC, {Maeda} K, {Nomoto} K. 2014  {Detection of
  the Gravitational Lens Magnifying a Type Ia Supernova}. {\em Science}
  \textbf{344}, 396--399.
(\href{http://dx.doi.org/10.1126/science.1250903}{10.1126/science.1250903})

\bibitem{2017Sci...356..291G}
{Goobar} A, {Amanullah} R, {Kulkarni} SR, {Nugent} PE, {Johansson} J, {Steidel}
  C, {Law} D, {M{\"o}rtsell} E, {Quimby} R, {Blagorodnova} N, {Brandeker} A,
  {Cao} Y, {Cooray} A, {Ferretti} R, {Fremling} C, {Hangard} L, {Kasliwal} M,
  {Kupfer} T, {Lunnan} R, {Masci} F, {Miller} AA, {Nayyeri} H, {Neill} JD,
  {Ofek} EO, {Papadogiannakis} S, {Petrushevska} T, {Ravi} V, {Sollerman} J,
  {Sullivan} M, {Taddia} F, {Walters} R, {Wilson} D, {Yan} L, {Yaron} O. 2017
  {iPTF16geu: A multiply imaged, gravitationally lensed type Ia supernova}.
  {\em Science} \textbf{356}, 291--295.
(\href{http://dx.doi.org/10.1126/science.aal2729}{10.1126/science.aal2729})

\bibitem{2023NatAs...7.1098G}
{Goobar} A, {Johansson} J, {Schulze} S, {Arendse} N, {Carracedo} AS, {Dhawan}
  S, {M{\"o}rtsell} E, {Fremling} C, {Yan} L, {Perley} D, {Sollerman} J,
  {Joseph} R, {Hinds} KR, {Meynardie} W, {Andreoni} I, {Bellm} E, {Bloom} J,
  {Collett} TE, {Drake} A, {Graham} M, {Kasliwal} M, {Kulkarni} SR, {Lemon} C,
  {Miller} AA, {Neill} JD, {Nordin} J, {Pierel} J, {Richard} J, {Riddle} R,
  {Rigault} M, {Rusholme} B, {Sharma} Y, {Stein} R, {Stewart} G, {Townsend} A,
  {Vinko} J, {Wheeler} JC, {Wold} A. 2023  {Uncovering a population of
  gravitational lens galaxies with magnified standard candle SN Zwicky}. {\em
  Nature Astronomy} \textbf{7}, 1098--1107.
(\href{http://dx.doi.org/10.1038/s41550-023-01981-3}{10.1038/s41550-023-01981-3})

\bibitem{2024SSRv..220...13S}
{Suyu} SH, {Goobar} A, {Collett} T, {More} A, {Vernardos} G. 2024  {Strong
  Gravitational Lensing and Microlensing of Supernovae}. {\em \ssr}
  \textbf{220}, 13.
(\href{http://dx.doi.org/10.1007/s11214-024-01044-7}{10.1007/s11214-024-01044-7})

\bibitem{2023ApJ...948..115P}
{Pierel} JDR, {Arendse} N, {Ertl} S, {Huang} X, {Moustakas} LA, {Schuldt} S,
  {Shajib} AJ, {Shu} Y, {Birrer} S, {Bronikowski} M, {Hjorth} J, {Suyu} SH,
  {Agarwal} S, {Agnello} A, {Bolton} AS, {Chakrabarti} S, {Cold} C, {Courbin}
  F, {Della Costa} JM, {Dhawan} S, {Engesser} M, {Fox} OD, {Gall} C, {Gomez} S,
  {Goobar} A, {Jha} SW, {Jimenez} C, {Johansson} J, {Larison} C, {Li} G,
  {Marques-Chaves} R, {Mao} S, {Mazzali} PA, {Perez-Fournon} I, {Petrushevska}
  T, {Poidevin} F, {Rest} A, {Sheu} W, {Shirley} R, {Silver} E, {Storfer} C,
  {Strolger} LG, {Treu} T, {Wojtak} R, {Zenati} Y. 2023  {LensWatch. I.
  Resolved HST Observations and Constraints on the Strongly Lensed Type Ia
  Supernova 2022qmx (``SN Zwicky'')}. {\em \apj} \textbf{948}, 115.
(\href{http://dx.doi.org/10.3847/1538-4357/acc7a6}{10.3847/1538-4357/acc7a6})

\bibitem{2020MNRAS.496.3270M}
{M{\"o}rtsell} E, {Johansson} J, {Dhawan} S, {Goobar} A, {Amanullah} R,
  {Goldstein} DA. 2020  {Lens modelling of the strongly lensed Type Ia
  supernova iPTF16geu}. {\em \mnras} \textbf{496}, 3270--3280.
(\href{http://dx.doi.org/10.1093/mnras/staa1600}{10.1093/mnras/staa1600})

\bibitem{2022A&A...662A..34D}
{Diego} JM, {Bernstein} G, {Chen} W, {Goobar} A, {Johansson} JP, {Kelly} PL,
  {M{\"o}rtsell} E, {Nightingale} JW. 2022  {Microlensing and the type Ia
  supernova iPTF16geu}. {\em \aap} \textbf{662}, A34.
(\href{http://dx.doi.org/10.1051/0004-6361/202143009}{10.1051/0004-6361/202143009})

\bibitem{2020MNRAS.491.2639D}
{Dhawan} S, {Johansson} J, {Goobar} A, {Amanullah} R, {M{\"o}rtsell} E, {Cenko}
  SB, {Cooray} A, {Fox} O, {Goldstein} D, {Kalender} R, {Kasliwal} M,
  {Kulkarni} SR, {Lee} WH, {Nayyeri} H, {Nugent} P, {Ofek} E, {Quimby} R. 2020
  {Magnification, dust, and time-delay constraints from the first resolved
  strongly lensed Type Ia supernova iPTF16geu}. {\em \mnras} \textbf{491},
  2639--2654.
(\href{http://dx.doi.org/10.1093/mnras/stz2965}{10.1093/mnras/stz2965})

\bibitem{2019ApJS..243....6G}
{Goldstein} DA, {Nugent} PE, {Goobar} A. 2019  {Rates and Properties of
  Supernovae Strongly Gravitationally Lensed by Elliptical Galaxies in
  Time-domain Imaging Surveys}. {\em \apjs} \textbf{243}, 6.
(\href{http://dx.doi.org/10.3847/1538-4365/ab1fe0}{10.3847/1538-4365/ab1fe0})

\bibitem{2024arXiv240600052S}
{Sagu{\'e}s Carracedo} A, {Goobar} A, {M{\"o}rtsell} E, {Arendse} N,
  {Johansson} J, {Townsend} A, {Dhawan} S, {Nordin} J, {Sollerman} J, {Schulze}
  S. 2024  {Detectability and Characterisation of Strongly Lensed Supernova
  Lightcurves in the Zwicky Transient Facility}. {\em arXiv e-prints} p.
  arXiv:2406.00052.

\bibitem{2024arXiv240806217A}
{Acevedo Barroso} JA, {O'Riordan} CM, {Cl{\'e}ment} B, {Tortora} C, {Collett}
  TE, {Courbin} F, {Gavazzi} R, {Metcalf} RB, {Busillo} V, {Andika} IT,
  {Cabanac} R, {Courtois} HM, {Crook-Mansour} J, {Delchambre} L, {Despali} G,
  {Ecker} LR, {Franco} A, {Holloway} P, {Jackson} N, {Jahnke} K, {Mahler} G,
  {Marchetti} L, {Matavulj} P, {Melo} A, {Meneghetti} M, {Moustakas} LA,
  {M{\"u}ller} O, {Nucita} AA, {Paulino-Afonso} A, {Pearson} J, {Rojas} K,
  {Scarlata} C, {Schuldt} S, {Serjeant} S, {Sluse} D, {Suyu} SH, {Vaccari} M,
  {Verma} A, {Vernardos} G, {Walmsley} M, {Bouy} H, {Walth} GL, {Powell} DM,
  {Bolzonella} M, {Cuillandre} JC, {Kluge} M, {Saifollahi} T, {Schirmer} M,
  {Stone} C, {Acebron} A, {Bazzanini} L, {D{\'\i}az-S{\'a}nchez} A, {Hogg} NB,
  {Koopmans} LVE, {Kruk} S, {Leuzzi} L, {Manj{\'o}n-Garc{\'\i}a} A, {Mannucci}
  F, {Nagam} BC, {Pearce-Casey} R, {Scharr{\'e}} L, {Wilde} J, {Altieri} B,
  {Amara} A, {Andreon} S, {Auricchio} N, {Baccigalupi} C, {Baldi} M, {Balestra}
  A, {Bardelli} S, {Basset} A, {Battaglia} P, {Bender} R, {Bonino} D,
  {Branchini} E, {Brescia} M, {Brinchmann} J, {Caillat} A, {Camera} S,
  {Candini} GP, {Capobianco} V, {Carbone} C, {Carretero} J, {Casas} S,
  {Castellano} M, {Castignani} G, {Cavuoti} S, {Cimatti} A, {Colodro-Conde} C,
  {Congedo} G, {Conselice} CJ, {Conversi} L, {Copin} Y, {Corcione} L, {Cropper}
  M, {Da Silva} A, {Degaudenzi} H, {De Lucia} G, {Dinis} J, {Dubath} F, {Dupac}
  X, {Dusini} S, {Farina} M, {Farrens} S, {Ferriol} S, {Frailis} M,
  {Franceschi} E, {Galeotta} S, {Garilli} B, {George} K, {Gillard} W, {Gillis}
  B, {Giocoli} C, {G{\'o}mez-Alvarez} P, {Grazian} A, {Grupp} F, {Guzzo} L,
  {Haugan} SVH, {Hoekstra} H, {Holmes} W, {Hook} I, {Hormuth} F, {Hornstrup} A,
  {Jhabvala} M, {Joachimi} B, {Keih{\"a}nen} E, {Kermiche} S, {Kiessling} A,
  {Kubik} B, {Kunz} M, {Kurki-Suonio} H, {Le Mignant} D, {Ligori} S, {Lilje}
  PB, {Lindholm} V, {Lloro} I, {Mainetti} G, {Maiorano} E, {Mansutti} O,
  {Marcin} S, {Marggraf} O, {Martinelli} M, {Martinet} N, {Marulli} F, {Massey}
  R, {Medinaceli} E, {Melchior} M, {Mellier} Y, {Merlin} E, {Meylan} G,
  {Moresco} M, {Moscardini} L, {Munari} E, {Nakajima} R, {Neissner} C, {Nichol}
  RC, {Niemi} SM, {Nightingale} JW, {Padilla} C, {Paltani} S, {Pasian} F,
  {Pedersen} K, {Percival} WJ, {Pettorino} V, {Pires} S, {Polenta} G, {Poncet}
  M, {Popa} LA, {Pozzetti} L, {Raison} F, {Rebolo} R, {Renzi} A, {Rhodes} J,
  {Riccio} G, {Romelli} E, {Roncarelli} M, {Rossetti} E, {Saglia} R, {Sakr} Z,
  {S{\'a}nchez} AG, {Sapone} D, {Schneider} P, {Schrabback} T, {Secroun} A,
  {Seidel} G, {Serrano} S, {Sirignano} C, {Sirri} G, {Skottfelt} J, {Stanco} L,
  {Steinwagner} J, {Tallada-Cresp{\'\i}} P, {Tavagnacco} D, {Taylor} AN,
  {Tereno} I, {Toledo-Moreo} R, {Torradeflot} F, {Tutusaus} I, {Valentijn} EA,
  {Valenziano} L, {Vassallo} T, {Wang} Y, {Weller} J, {Zucca} E, {Burigana} C,
  {Scottez} V, {Viel} M. 2024  {Euclid: The Early Release Observations Lens
  Search Experiment}. {\em arXiv e-prints} p. arXiv:2408.06217.
(\href{http://dx.doi.org/10.48550/arXiv.2408.06217}{10.48550/arXiv.2408.06217})

\bibitem{Bellm+2019}
{Bellm} EC, {Kulkarni} SR, {Graham} MJ, {Dekany} R, {Smith} RM, {Riddle} R,
  {Masci} FJ, {Helou} G, {Prince} TA, {Adams} SM, {Barbarino} C, {Barlow} T,
  {Bauer} J, {Beck} R, {Belicki} J, {Biswas} R, {Blagorodnova} N, {Bodewits} D,
  {Bolin} B, {Brinnel} V, {Brooke} T, {Bue} B, {Bulla} M, {Burruss} R, {Cenko}
  SB, {Chang} CK, {Connolly} A, {Coughlin} M, {Cromer} J, {Cunningham} V, {De}
  K, {Delacroix} A, {Desai} V, {Duev} DA, {Eadie} G, {Farnham} TL, {Feeney} M,
  {Feindt} U, {Flynn} D, {Franckowiak} A, {Frederick} S, {Fremling} C,
  {Gal-Yam} A, {Gezari} S, {Giomi} M, {Goldstein} DA, {Golkhou} VZ, {Goobar} A,
  {Groom} S, {Hacopians} E, {Hale} D, {Henning} J, {Ho} AYQ, {Hover} D,
  {Howell} J, {Hung} T, {Huppenkothen} D, {Imel} D, {Ip} WH, {Ivezi{\'c}}
  {\v{Z}}, {Jackson} E, {Jones} L, {Juric} M, {Kasliwal} MM, {Kaspi} S, {Kaye}
  S, {Kelley} MSP, {Kowalski} M, {Kramer} E, {Kupfer} T, {Landry} W, {Laher}
  RR, {Lee} CD, {Lin} HW, {Lin} ZY, {Lunnan} R, {Giomi} M, {Mahabal} A, {Mao}
  P, {Miller} AA, {Monkewitz} S, {Murphy} P, {Ngeow} CC, {Nordin} J, {Nugent}
  P, {Ofek} E, {Patterson} MT, {Penprase} B, {Porter} M, {Rauch} L,
  {Rebbapragada} U, {Reiley} D, {Rigault} M, {Rodriguez} H, {van Roestel} J,
  {Rusholme} B, {van Santen} J, {Schulze} S, {Shupe} DL, {Singer} LP,
  {Soumagnac} MT, {Stein} R, {Surace} J, {Sollerman} J, {Szkody} P, {Taddia} F,
  {Terek} S, {Van Sistine} A, {van Velzen} S, {Vestrand} WT, {Walters} R,
  {Ward} C, {Ye} QZ, {Yu} PC, {Yan} L, {Zolkower} J. 2019  {The Zwicky
  Transient Facility: System Overview, Performance, and First Results}. {\em
  \pasp} \textbf{131}, 018002.
(\href{http://dx.doi.org/10.1088/1538-3873/aaecbe}{10.1088/1538-3873/aaecbe})

\bibitem{2020ApJ...895...32F}
{Fremling} C, {Miller} AA, {Sharma} Y, {Dugas} A, {Perley} DA, {Taggart} K,
  {Sollerman} J, {Goobar} A, {Graham} ML, {Neill} JD, {Nordin} J, {Rigault} M,
  {Walters} R, {Andreoni} I, {Bagdasaryan} A, {Belicki} J, {Cannella} C,
  {Bellm} EC, {Cenko} SB, {De} K, {Dekany} R, {Frederick} S, {Golkhou} VZ,
  {Graham} MJ, {Helou} G, {Ho} AYQ, {Kasliwal} MM, {Kupfer} T, {Laher} RR,
  {Mahabal} A, {Masci} FJ, {Riddle} R, {Rusholme} B, {Schulze} S, {Shupe} DL,
  {Smith} RM, {van Velzen} S, {Yan} L, {Yao} Y, {Zhuang} Z, {Kulkarni} SR. 2020
   {The Zwicky Transient Facility Bright Transient Survey. I. Spectroscopic
  Classification and the Redshift Completeness of Local Galaxy Catalogs}. {\em
  \apj} \textbf{895}, 32.
(\href{http://dx.doi.org/10.3847/1538-4357/ab8943}{10.3847/1538-4357/ab8943})

\bibitem{2023MNRAS.526.4296S}
{Sainz de Murieta} A, {Collett} TE, {Magee} MR, {Weisenbach} L, {Krawczyk} CM,
  {Enzi} W. 2023  {Lensed Type Ia supernovae in light of SN Zwicky and
  iPTF16geu}. {\em \mnras} \textbf{526}, 4296--4307.
(\href{http://dx.doi.org/10.1093/mnras/stad3031}{10.1093/mnras/stad3031})

\bibitem{2024arXiv240518589T}
{Townsend} A, {Nordin} J, {Sagu{\'e}s Carracedo} A, {Kowalski} M, {Arendse} N,
  {Dhawan} S, {Goobar} A, {Johansson} J, {M{\"o}rtsell} E, {Schulze} S,
  {Andreoni} I, {Fern{\'a}ndez} E, {Kim} AG, {Nugent} PE, {Prada} F, {Rigault}
  M, {Sarin} N, {Sharma} D, {Bellm} EC, {Coughlin} MW, {Dekany} R, {Groom} SL,
  {Lacroix} L, {Laher} RR, {Riddle} R, {Aguilar} J, {Ahlen} S, {Bailey} S,
  {Brooks} D, {Claybaugh} T, {de la Macorra} A, {Dey} A, {Dey} B, {Doel} P,
  {Fanning} K, {Forero-Romero} JE, {Gazta{\~n}aga} E, {Gontcho} SGA,
  {Honscheid} K, {Howlett} C, {Kisner} T, {Kremin} A, {Lambert} A, {Landriau}
  M, {Le Guillou} L, {Levi} ME, {Manera} M, {Meisner} A, {Miquel} R,
  {Moustakas} J, {Mueller} E, {Myers} AD, {Nie} J, {Palanque-Delabrouille} N,
  {Poppett} C, {Rezaie} M, {Rossi} G, {Sanchez} E, {Schlegel} D, {Schubnell} M,
  {Seo} H, {Sprayberry} D, {Tarl{\'e}} G, {Zou} H. 2024  {Candidate
  strongly-lensed Type Ia supernovae in the Zwicky Transient Facility archive}.
  {\em arXiv e-prints} p. arXiv:2405.18589.
(\href{http://dx.doi.org/10.48550/arXiv.2405.18589}{10.48550/arXiv.2405.18589})

\bibitem{2023ARNPS..73..153K}
{Kamionkowski} M, {Riess} AG. 2023  {The Hubble Tension and Early Dark Energy}.
  {\em Annual Review of Nuclear and Particle Science} \textbf{73}, 153--180.
(\href{http://dx.doi.org/10.1146/annurev-nucl-111422-024107}{10.1146/annurev-nucl-111422-024107})

\bibitem{2022arXiv221010833B}
{Birrer} S, {Millon} M, {Sluse} D, {Shajib} AJ, {Courbin} F, {Koopmans} LVE,
  {Suyu} SH, {Treu} T. 2022  {Time-Delay Cosmography: Measuring the Hubble
  Constant and other cosmological parameters with strong gravitational
  lensing}. {\em arXiv e-prints} p. arXiv:2210.10833.
(\href{http://dx.doi.org/10.48550/arXiv.2210.10833}{10.48550/arXiv.2210.10833})

\bibitem{Falco+1985}
{Falco} EE, {Gorenstein} MV, {Shapiro} II. 1985  {On model-dependent bounds on
  H 0 from gravitational images : application to Q 0957+561 A, B.}. {\em \apjl}
  \textbf{289}, L1--L4.
(\href{http://dx.doi.org/10.1086/184422}{10.1086/184422})

\bibitem{Birrer+2021}
{Birrer} S, {Dhawan} S, {Shajib} AJ. 2021  {The Hubble constant from strongly
  lensed supernovae with standardizable magnifications}. {\em arXiv e-prints}
  p. arXiv:2107.12385.

\bibitem{Rodney+2016}
{Rodney} SA, {Strolger} LG, {Kelly} PL, {Brada{\v{c}}} M, {Brammer} G,
  {Filippenko} AV, {Foley} RJ, {Graur} O, {Hjorth} J, {Jha} SW, {McCully} C,
  {Molino} A, {Riess} AG, {Schmidt} KB, {Selsing} J, {Sharon} K, {Treu} T,
  {Weiner} BJ, {Zitrin} A. 2016  {SN Refsdal: Photometry and Time Delay
  Measurements of the First Einstein Cross Supernova}. {\em \apj} \textbf{820},
  50.
(\href{http://dx.doi.org/10.3847/0004-637X/820/1/50}{10.3847/0004-637X/820/1/50})

\bibitem{Kelly+2023a}
{Kelly} PL, {Rodney} S, {Treu} T, {Oguri} M, {Chen} W, {Zitrin} A, {Birrer} S,
  {Bonvin} V, {Dessart} L, {Diego} JM, {Filippenko} AV, {Foley} RJ, {Gilman} D,
  {Hjorth} J, {Jauzac} M, {Mandel} K, {Millon} M, {Pierel} J, {Sharon} K,
  {Thorp} S, {Williams} L, {Broadhurst} T, {Dressler} A, {Graur} O, {Jha} S,
  {McCully} C, {Postman} M, {Schmidt} KB, {Tucker} BE, {von der Linden} A. 2023
   {Constraints on the Hubble constant from supernova Refsdal's reappearance}.
  {\em Science} \textbf{380}, abh1322.
(\href{http://dx.doi.org/10.1126/science.abh1322}{10.1126/science.abh1322})

\bibitem{2024A&A...684L..23G}
{Grillo} C, {Pagano} L, {Rosati} P, {Suyu} SH. 2024  {Cosmography with
  supernova Refsdal through time-delay cluster lensing: Independent
  measurements of the Hubble constant and geometry of the Universe}. {\em \aap}
  \textbf{684}, L23.
(\href{http://dx.doi.org/10.1051/0004-6361/202449278}{10.1051/0004-6361/202449278})

\bibitem{2024arXiv240319029C}
{Chen} W, {Kelly} PL, {Frye} BL, {Pierel} J, {Willner} SP, {Pascale} M, {Cohen}
  SH, {Conselice} CJ, {Engesser} M, {Furtak} LJ, {Gilman} D, {Grogin} NA,
  {Huber} S, {Jha} SW, {Johansson} J, {Koekemoer} AM, {Larison} C, {Meena} AK,
  {Siebert} MR, {Windhorst} RA, {Yan} H, {Zitrin} A. 2024  {JWST Spectroscopy
  of SN H0pe: Classification and Time Delays of a Triply-imaged Type Ia
  Supernova at z = 1.78}. {\em arXiv e-prints} p. arXiv:2403.19029.
(\href{http://dx.doi.org/10.48550/arXiv.2403.19029}{10.48550/arXiv.2403.19029})

\bibitem{2024arXiv240318954P}
{Pierel} JDR, {Frye} BL, {Pascale} M, {Caminha} GB, {Chen} W, {Dhawan} S,
  {Gilman} D, {Grayling} M, {Huber} S, {Kelly} P, {Thorp} S, {Arendse} N,
  {Birrer} S, {Bronikowski} M, {Canameras} R, {Coe} D, {Cohen} SH, {Conselice}
  CJ, {Driver} SP, {Dsilva} JCJ, {Engesser} M, {Foo} N, {Gall} C, {Garuda} N,
  {Grillo} C, {Grogin} NA, {Henderson} J, {Hjorth} J, {Jansen} RA, {Johansson}
  J, {Kamieneski} PS, {Koekemoer} AM, {Larison} C, {Marshall} MA, {Moustakas}
  LA, {Nonino} M, {Ortiz}, R. I, {Petrushevska} T, {Pirzkal} N, {Robotham} A,
  {Ryan}, R.~E. J, {Schuldt} S, {Strolger} LG, {Summers} J, {Suyu} SH, {Treu}
  T, {Willmer} CNA, {Windhorst} RA, {Yan} H, {Zitrin} A, {Acebron} A,
  {Chakrabarti} S, {Coulter} DA, {Fox} OD, {Huang} X, {Jha} SW, {Li} G,
  {Mazzali} PA, {Meena} AK, {Perez-Fournon} I, {Poidevin} F, {Rest} A, {Riess}
  AG. 2024  {JWST Photometric Time-Delay and Magnification Measurements for the
  Triply-Imaged Type Ia ``Supernova H0pe'' at z = 1.78}. {\em arXiv e-prints}
  p. arXiv:2403.18954.
(\href{http://dx.doi.org/10.48550/arXiv.2403.18954}{10.48550/arXiv.2403.18954})

\bibitem{2024arXiv240318902P}
{Pascale} M, {Frye} BL, {Pierel} JDR, {Chen} W, {Kelly} PL, {Cohen} SH,
  {Windhorst} RA, {Riess} AG, {Kamieneski} PS, {Diego} JM, {Meena} AK, {Cha} S,
  {Oguri} M, {Zitrin} A, {Jee} MJ, {Foo} N, {Leimbach} R, {Koekemoer} AM,
  {Conselice} CJ, {Dai} L, {Goobar} A, {Siebert} MR, {Strolger} L, {Willner}
  SP. 2024  {SN H0pe: The First Measurement of $H_0$ from a Multiply-Imaged
  Type Ia Supernova, Discovered by JWST}. {\em arXiv e-prints} p.
  arXiv:2403.18902.
(\href{http://dx.doi.org/10.48550/arXiv.2403.18902}{10.48550/arXiv.2403.18902})

\bibitem{2006ApJ...649..939K}
{Kasen} D. 2006  {Secondary Maximum in the Near-Infrared Light Curves of Type
  Ia Supernovae}. {\em \apj} \textbf{649}, 939--953.
(\href{http://dx.doi.org/10.1086/506588}{10.1086/506588})

\bibitem{2014MNRAS.441.3249D}
{Dessart} L, {Hillier} DJ, {Blondin} S, {Khokhlov} A. 2014  {Critical
  ingredients of Type Ia supernova radiative-transfer modelling}. {\em \mnras}
  \textbf{441}, 3249--3270.
(\href{http://dx.doi.org/10.1093/mnras/stu789}{10.1093/mnras/stu789})

\bibitem{2015MNRAS.449.3581J}
{Jack} D, {Baron} E, {Hauschildt} PH. 2015  {Identification of the feature that
  causes the I-band secondary maximum of a Type Ia supernova}. {\em \mnras}
  \textbf{449}, 3581--3586.
(\href{http://dx.doi.org/10.1093/mnras/stv474}{10.1093/mnras/stv474})

\bibitem{hsiao2007}
{Hsiao} EY, {Conley} A, {Howell} DA, {Sullivan} M, {Pritchet} CJ, {Carlberg}
  RG, {Nugent} PE, {Phillips} MM. 2007  {K-Corrections and Spectral Templates
  of Type Ia Supernovae}. {\em \apj} \textbf{663}, 1187--1200.
(\href{http://dx.doi.org/10.1086/518232}{10.1086/518232})

\bibitem{Johansson+2021}
{Johansson} J, {Goobar} A, {Price} SH, {Sagu{\'e}s Carracedo} A, {Della Bruna}
  L, {Nugent} PE, {Dhawan} S, {M{\"o}rtsell} E, {Papadogiannakis} S,
  {Amanullah} R, {Goldstein} D, {Cenko} SB, {De} K, {Dugas} A, {Kasliwal} MM,
  {Kulkarni} SR, {Lunnan} R. 2021  {Spectroscopy of the first resolved strongly
  lensed Type Ia supernova iPTF16geu}. {\em \mnras} \textbf{502}, 510--520.
(\href{http://dx.doi.org/10.1093/mnras/staa3829}{10.1093/mnras/staa3829})

\bibitem{Pierel2019}
{Pierel} JR, {Rodney} SA. 2019  {SNTD: Supernova Time Delays}. Astrophysics
  Source Code Library, record ascl:1902.001.

\bibitem{Guy:2007js}
Guy J, Astier P, Baumont S, Hardin D, Pain R, Regnault N, Basa S, Carlberg RG,
  Conley A, Fabbro S, Fouchez D, Hook IM, Howell DA, Perrett K, Pritchet CJ,
  Rich J, Sullivan M, Antilogus P, Aubourg E, Bazin G, Bronder J, Filiol M,
  Palanque-Delabrouille N, Ripoche P, Ruhlmann-Kleider V. 2007  {SALT2: using
  distant supernovae to improve the use of type Ia supernovae as distance
  indicators}. {\em \aap} \textbf{466}, 11--21.

\bibitem{2015MNRAS.453.3300A}
{Amanullah} R, {Johansson} J, {Goobar} A, {Ferretti} R, {Papadogiannakis} S,
  {Petrushevska} T, {Brown} PJ, {Cao} Y, {Contreras} C, {Dahle} H, {Elias-Rosa}
  N, {Fynbo} JPU, {Gorosabel} J, {Guaita} L, {Hangard} L, {Howell} DA, {Hsiao}
  EY, {Kankare} E, {Kasliwal} M, {Leloudas} G, {Lundqvist} P, {Mattila} S,
  {Nugent} P, {Phillips} MM, {Sandberg} A, {Stanishev} V, {Sullivan} M,
  {Taddia} F, {{\"O}stlin} G, {Asadi} S, {Herrero-Illana} R, {Jensen} JJ,
  {Karhunen} K, {Lazarevic} S, {Varenius} E, {Santos} P, {Sridhar} SS,
  {Wallstr{\"o}m} SHJ, {Wiegert} J. 2015  {Diversity in extinction laws of Type
  Ia supernovae measured between 0.2 and 2 {\ensuremath{\mu}}m}. {\em \mnras}
  \textbf{453}, 3300--3328.
(\href{http://dx.doi.org/10.1093/mnras/stv1505}{10.1093/mnras/stv1505})

\bibitem{More+2017}
{More} A, {Suyu} SH, {Oguri} M, {More} S, {Lee} CH. 2017  {Interpreting the
  Strongly Lensed Supernova iPTF16geu: Time Delay Predictions, Microlensing,
  and Lensing Rates}. {\em \apjl} \textbf{835}, L25.
(\href{http://dx.doi.org/10.3847/2041-8213/835/2/L25}{10.3847/2041-8213/835/2/L25})

\bibitem{DoblerKeeton2006}
{Dobler} G, {Keeton} CR. 2006  {Microlensing of Lensed Supernovae}. {\em \apj}
  \textbf{653}, 1391--1399.
(\href{http://dx.doi.org/10.1086/508769}{10.1086/508769})

\bibitem{Foxley-Marrable+2018}
{Foxley-Marrable} M, {Collett} TE, {Vernardos} G, {Goldstein} DA, {Bacon} D.
  2018  {The impact of microlensing on the standardization of strongly lensed
  Type Ia supernovae}. {\em \mnras} \textbf{478}, 5081--5090.
(\href{http://dx.doi.org/10.1093/mnras/sty1346}{10.1093/mnras/sty1346})

\bibitem{2024arXiv240303264W}
{Weisenbach} L, {Collett} T, {Sainz de Murieta} A, {Krawczyk} C, {Vernardos} G,
  {Enzi} W, {Lundgren} A. 2024  {How to Break the Mass Sheet Degeneracy with
  the Lightcurves of Microlensed Type Ia Supernovae}. {\em arXiv e-prints} p.
  arXiv:2403.03264.
(\href{http://dx.doi.org/10.48550/arXiv.2403.03264}{10.48550/arXiv.2403.03264})

\bibitem{2024arXiv240608919S}
{Shajib} AJ, {Smith} GP, {Birrer} S, {Verma} A, {Arendse} N, {Collett} TE. 2024
   {Strong gravitational lenses from the Vera C. Rubin Observatory}. {\em arXiv
  e-prints} p. arXiv:2406.08919.
(\href{http://dx.doi.org/10.48550/arXiv.2406.08919}{10.48550/arXiv.2406.08919})

\bibitem{Wojtak+2019}
{Wojtak} R, {Hjorth} J, {Gall} C. 2019  {Magnified or multiply imaged? - Search
  strategies for gravitationally lensed supernovae in wide-field surveys}. {\em
  \mnras} \textbf{487}, 3342--3355.
(\href{http://dx.doi.org/10.1093/mnras/stz1516}{10.1093/mnras/stz1516})

\bibitem{2024arXiv240704080S}
{Sainz de Murieta} A, {Collett} TE, {Magee} MR, {Pierel} JDR, {Enzi} WJR,
  {Lokken} M, {Gagliano} A, {Ryczanowski} D. 2024  {Find the haystacks, then
  look for needles: The rate of strongly lensed transients in galaxy-galaxy
  strong gravitational lenses}. {\em arXiv e-prints} p. arXiv:2407.04080.
(\href{http://dx.doi.org/10.48550/arXiv.2407.04080}{10.48550/arXiv.2407.04080})

\bibitem{2024arXiv240601434D}
{Dhawan} S, {Mortsell} E, {Johansson} J, {Goobar} A, {Rigault} M, {Smith} M,
  {Maguire} K, {Nordin} J, {Dimitriadis} G, {Nugent} PE, {Galbany} L,
  {Sollerman} J, {de Jaeger} T, {Terwel} JH, {Kim} YL, {Burgaz} U, {Helou} G,
  {Purdum} J, {Groom} SL, {Laher} R, {Healy} B. 2024  {ZTF SN Ia DR2:
  Cosmology-independent constraints on Type Ia supernova standardisation from
  supernova siblings}. {\em arXiv e-prints} p. arXiv:2406.01434.
(\href{http://dx.doi.org/10.48550/arXiv.2406.01434}{10.48550/arXiv.2406.01434})

\bibitem{2020ApJ...896L..13S}
{Scolnic} D, {Smith} M, {Massiah} A, {Wiseman} P, {Brout} D, {Kessler} R,
  {Davis} TM, {Foley} RJ, {Galbany} L, {Hinton} SR, {Hounsell} R, {Kelsey} L,
  {Lidman} C, {Macaulay} E, {Morgan} R, {Nichol} RC, {M{\"o}ller} A, {Popovic}
  B, {Sako} M, {Sullivan} M, {Thomas} BP, {Tucker} BE, {Abbott} TMC, {Aguena}
  M, {Allam} S, {Annis} J, {Avila} S, {Bechtol} K, {Bertin} E, {Brooks} D,
  {Burke} DL, {Rosell} AC, {Carollo} D, {Kind} MC, {Carretero} J, {Costanzi} M,
  {da Costa} LN, {De Vicente} J, {Desai} S, {Diehl} HT, {Doel} P,
  {Drlica-Wagner} A, {Eckert} K, {Eifler} TF, {Everett} S, {Flaugher} B,
  {Fosalba} P, {Frieman} J, {Garc{\'\i}a-Bellido} J, {Gaztanaga} E, {Gerdes}
  DW, {Glazebrook} K, {Gruen} D, {Gruendl} RA, {Gschwend} J, {Gutierrez} G,
  {Hartley} WG, {Hollowood} DL, {Honscheid} K, {James} DJ, {Kuehn} K,
  {Kuropatkin} N, {Lewis} GF, {Li} TS, {Lima} M, {Maia} MAG, {Marshall} JL,
  {Menanteau} F, {Miquel} R, {Palmese} A, {Paz-Chinch{\'o}n} F, {Plazas} AA,
  {Pursiainen} M, {Sanchez} E, {Scarpine} V, {Schubnell} M, {Serrano} S,
  {Sevilla-Noarbe} I, {Sommer} NE, {Suchyta} E, {Swanson} MEC, {Tarle} G,
  {Varga} TN, {Walker} AR, {Wilkinson} R, {DES Collaboration}. 2020  {Supernova
  Siblings: Assessing the Consistency of Properties of Type Ia Supernovae that
  Share the Same Parent Galaxies}. {\em \apjl} \textbf{896}, L13.
(\href{http://dx.doi.org/10.3847/2041-8213/ab8735}{10.3847/2041-8213/ab8735})

\bibitem{2024MNRAS.531.3509A}
{Arendse} N, {Dhawan} S, {Sagu{\'e}s Carracedo} A, {Peiris} HV, {Goobar} A,
  {Wojtak} R, {Alves} C, {Biswas} R, {Huber} S, {Birrer} S, {The LSST Dark
  Energy Science Collaboration}. 2024  {Detecting strongly lensed type Ia
  supernovae with LSST}. {\em \mnras} \textbf{531}, 3509--3523.
(\href{http://dx.doi.org/10.1093/mnras/stae1356}{10.1093/mnras/stae1356})

\end{thebibliography}






\end{document}